\documentclass[preprint,12pt]{elsarticle}

\usepackage{amssymb}

\begin{document}

\begin{frontmatter}



\title{Two-dimensional topological insulators \\ in quantizing magnetic fields}


\author{G. Tkachov and E. M. Hankiewicz}

\address{Institute for Theoretical Physics und Astrophysics, University of W\"urzburg, Germany}

\begin{abstract}
Two-dimensional topological insulators are characterized by gapped bulk states and 
gapless helical edge states, i.e. time-reversal symmetric edge states accommodating a pair of counter-propagating electrons. 
An external magnetic field breaks the time-reversal symmetry. What happens to the edge states in this case? 
In this paper we analyze the edge-state spectrum and longitudinal conductance in a two-dimensional topological insulator subject to a quantizing magnetic field. 
We show that the helical edge states exist also in this case. 
The strong magnetic field modifies the group velocities of the counter-propagating channels which are no longer identical. 
The helical edge states with different group velocities are particularly  prone to get coupled via backscattering, 
which leads to the suppression of the longitudinal edge magnetoconductance.  
\end{abstract}


\end{frontmatter}


\section{Introduction}
\label{intro}

The discovery of two-dimensional (2D)~\cite{Kane05,Bernevig06,Koenig07} and three-dimensional (3D)~\cite{Fu07,Moore07,Hsieh08,Xia09,Chen09} topological insulator phases in materials with strong spin-orbit coupling has stimulated vigorous research in this field~\cite{Koenig08,Hasan10,Qi10}.  
Topological insulators (TIs) are distinct from ordinary band insulators and semimetals by the presence of protected surface (in 3D) or edge (in 2D) states. In a 3D TI the surface state has a Dirac-cone spectrum with zero band gap as a consequence of time-reversal symmetry (TRS). If TRS is broken, an energy gap is induced at the Dirac point,
and the surface state exhibits the quantum Hall (QH) effect~\cite{Qi08,Essin09,Tse10a,Tse10b,Bruene11} and rich magneto-electric phenomena~\cite{Qi08,Essin09,Tse10a,Tse10b,Maciejko10a,Garate10,GT10c} related to axion electrodynamics~\cite{Wilczek87}. 

The 2D TIs have been realized in HgTe/CdTe quantum wells (QWs)~\cite{Bernevig06,Koenig07,Roth09}. Their electronic bands 
form a single double-degenerate Dirac valley~\cite{Buettner11}. The double degeneracy of the QW bands allows for an energy gap at the Dirac point without TRS breaking, so that the conduction electrons mimic the behaviour of massive Dirac fermions with specific mobility~\cite{GT11a} and weak antilocalization effects~\cite{GT11b}.
The unique feature of the HgTe/CdTe QWs is that their band gap inverts its sign upon changing the thickness of the HgTe layer~\cite{Koenig07}. The topologically nontrivial phase - the quantum spin Hall (QSH) state~\cite{Kane05,Bernevig06,Koenig07,Roth09} - occurs when the Fermi level lies within the inverted band gap and is characterized by gapless quasi-one-dimensional states on sample edges, while the states in the 2D bulk are fully gapped.
Unlike the chiral QH edge states~\cite{Halperin82,MacDonald84} the edge modes of a QSH insulator possess the TRS because they accommodate 
counter-propagating opposite-spin electrons and, for this reason, are frequently, called helical.  
One particular consequence of the TRS is that the counter-propagating helical channels have the same group velocities.

In experiments on HgTe/CdTe QWs~\cite{Koenig07}, the QSH state
was detected by measuring the longitudinal electric conductance of two spin channels 
propagating in the same direction on opposite edges of the sample.
This finding was further substantiated by the observed suppression of 
the edge transport in an external magnetic field~\cite{Koenig07},
which is expected since the magnetic breaks the TRS. 
However, the concrete scenario of the TRS breaking may depend on a number of poorly controlled factors such as the degree of bulk-inversion asymmetry, 
strength and type of disorder~\cite{Maciejko10b,GT10a}, which require further investigations.   

\begin{figure}[t!]
\begin{center}
\resizebox{0.4\columnwidth}{!}{%
  \includegraphics{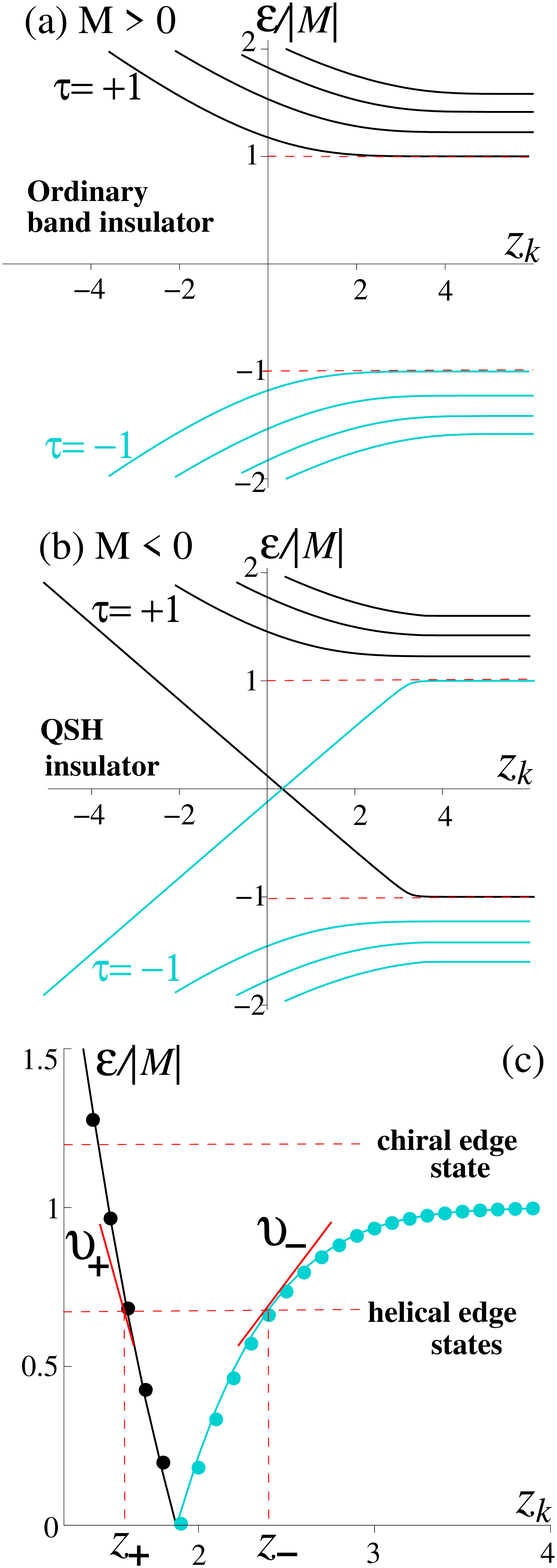}
}
\end{center}
\caption{
Edge-state energy as a function of center-of-oscillator coordinate
$z_k=-2\lambda\, k\, {\rm sgn} (eB)$ where $k$ is the wave vector along the edge:
(a) ordinary band insulator and (b) QSH insulator with two counterpropagating (helical) gapless states  
merging with flat bulk Landau levels at large $z_k$; $\hbar\upsilon/\lambda |M| =0.5$ and $eB>0$. 
(c) Edge states in a strong magnetic field ($\hbar\upsilon/\lambda |M|\geq 10$). 
Helical spectrum turns chiral at band gap energy $\epsilon=|M|$. 
Points indicate numerical solution of Eq. (\ref{Zeros}), whereas 
solid lines are the analytical result, Eq. (\ref{E_strong}). 
}
\label{Edge}
\end{figure}

In this paper we characterize the TRS breaking in the QSH regime in terms of the modification of the edge-state dispersion in a magnetic field. 
We demonstrate that the counter-propagating helical edge states persist in a strong quantizing magnetic field due to the fact that they are protected by the band gap. 
However, they have now distinctly different group velocities: One of the edge modes merges with the bulk lowest Landau level and therefore becomes slower than the other (see also Fig. \ref{Edge}).      
At the band gap energy (corresponding to the bulk lowest Landau level) the edge spectrum changes from helical to chiral.
Such a transformation occurs as the Fermi level is driven   
from the band gap into the Landau-quantized conduction or valence band 
where a dissipationless QH state sets in. 
We find that in the QSH regime (i.e. below the band gap) the ``slow'' and ``fast'' edge modes  
are prone to get coupled by weak disorder that generates backscattering between the counter-propagating channels. 
This leads to suppression of the two-terminal longitudinal edge conductance $g$ as a function of both Fermi energy and magnetic field $B$: 
\begin{equation}
g(\epsilon,B)\propto (|M| - |\epsilon|)^{2N}/B^{2N}, \quad |\epsilon| \to |M|.
\label{g_as}
\end{equation}
Here $\epsilon$ indicates the position of the Fermi level 
with respect to the band gap energy $|M|$. 
Equation (\ref{g_as}) contrasts the behaviour of the zero-field conductance 
which increases as the Fermi level is pushed 
into the metallic-type conduction or valence band~\cite{Koenig07,Roth09}. 
Also, unlike the exponential $B$ decay in strongly disordered systems~\cite{Maciejko10b}, 
Eq. (\ref{g_as}) describes a power-law magnetoconductance.   

Equation (\ref{g_as}) assumes the presence of a few ($N$) backscattering centers on the edge 
such as sample inhomogeneities where electronic trap states can interact with the edge channels 
randomizing their propagation directions~\cite{Roth09}.     
Although in a zero field this effect is believed to be weak, 
we show that near the band gap the backscattering is dramatically enhanced 
due to the flattening of the dispersion of one of the coupled QSH modes.
According to Eq.~(\ref{g_as}), the analysis of the power of the magnetoconductance decay   
can be a useful tool to determine the quality of the QSH devices.  

\section{ From helical to chiral edge states in a quantizing magnetic field }

\subsection{ Boundary problem and its solution }

We will first analyze the edge states in scattering-free HgTe QWs using 
the effective 4-band model derived in Ref.~\cite{Bernevig06}.
In this approach one works in the basis of the four states near the $\Gamma$  (${\bf k}=0$) 
point of the Brillouin zone:
$|e_1 +\rangle$, $|h_1 +\rangle$, $|e_1 -\rangle$, and $|h_1 -\rangle$,
where $e_1$ and $h_1$ are the s-like electron and p-like hole QW subbands, respectively.  
The index $\tau=\pm$ accounts for the spin degree of freedom.
The effective 2D Hamiltonian can be approximated by a diagonal matrix in ${\tau}$ space~\cite{Bernevig06};
\begin{eqnarray}
H=
\left(
  \begin{array}{cc}
    h_{\bf k} & 0 \\
    0 & h^{\ast}_{\bf -k} \\
  \end{array}
\right),
h_{\bf k} = {\bf d}_{\bf k}\mbox{\boldmath$\sigma$},
\,
{\bf d}_{\bf k} =(\hbar\upsilon k_x, -\hbar\upsilon k_y, M).
\label{Heff}
\end{eqnarray}
where Pauli matrices $\sigma_{x,y,z}$ act in subband space,  
$\upsilon\approx 5.5 \times 10^{5} $ms$^{-1}$ is the effective velocity~\cite{Koenig08},
and $M$ determines the band gap $E_g=|M|$ at ${\bf k}=0$.
In Eq.~(\ref{Heff}) we omit the terms $\propto {\bf k}^2$, which are small near 
the $\Gamma$ point, and assume no coupling between the Kramers partners, 
which is a good approximation for symmetric HgTe quantum wells~\cite{Buettner11,Rothe10}. 
Up to a unitary transformation, Eq.~(\ref{Heff}) is equivalent to a massive Dirac Hamiltonian 
\begin{equation}
H_D = \hbar\upsilon \tau_z\mbox{\boldmath$\sigma$}{\bf k} + M\tau_z\sigma_z, 
\label{H_D}
\end{equation}
$\tau_z$ is the Pauli matrix in spin space.  
Following the previous studies of edge states in graphene~\cite{GT07,GT09a,GT09b,GT09c,TB_GF}
we will work with the matrix retarded Green's function defined by the equation  
\begin{eqnarray}
[\epsilon\, I  - H_D]{\hat G}({\bf r},{\bf r}^\prime)=I \delta( {\bf r}-{\bf r}^\prime ),
\label{Eq_G}
\end{eqnarray}
where
$
{\bf k}=-i{\bf\nabla}  - e{\bf A}({\bf r})/c\hbar,
$
${\bf A}({\bf r})=(-By,0,0)$ is the vector potential of an external magnetic field $B$, 
and $I=\tau_0\sigma_0={\rm diag}(1,1,1,1)$. 
Assuming a sufficiently wide sample, we find ${\hat G}({\bf r},{\bf r}^\prime)$ near one of the edges,
e.g. $y=0$, using the boundary condition 
\begin{eqnarray} 
{\hat G}({\bf r},{\bf r}^\prime)|_{y=0}=
\tau_0\sigma_x\, {\hat G}({\bf r},{\bf r}^\prime)|_{y=0},
\label{BC}
\end{eqnarray}
equivalent to confinement by infinite "mass" at $y<0$~\cite{Berry87}.
This boundary condition can be obtained by introducing a large mass term ($M \to \infty  $) 
outside the physical area of the system~\cite{Berry87}. 
Our results do not however strongly depend on the choice of the boundary condition 
since the origin of the QSH edge states is topological: 
a mass domain wall in the inverted regime with $M<0$ in the bulk~\cite{Bernevig06,Volkov85}. 

The matrix ${\hat G}={\rm diag}({\hat G}_+, {\hat G}_-)$ is diagonal in $\tau$ space,
and each ${\hat G}_\tau$ can be diagonalized in e,h space:
\begin{eqnarray}
{\hat G}_\tau &=& 
\left(
\begin{array}{cc}
G_{ee|\tau}  & G_{eh|\tau}\\
G_{he|\tau} & G_{hh|\tau}
\end{array}
\right)=
\label{G_eh}\\
&=&
\left(
\begin{array}{cc}
1  & \frac{\upsilon(p_x-ip_y)}{\tau\epsilon - M}\\
\frac{\upsilon(p_x+ip_y)}{\tau\epsilon + M} & 1
\end{array}
\right)
\left(
\begin{array}{cc}
G_{ee|\tau}  & 0\\
0 & G_{hh|\tau}
\end{array}
\right).
\label{G_diag}
\end{eqnarray}
Expanding ${\hat G}$ in plane waves ${\rm e}^{ikx}$ yields 
the boundary problem for the diagonal elements:
\begin{eqnarray}
&
\left[\partial^2_z - \frac{(z - z_k)^2}{4} - a \right]G_{ee|\tau k}=
\frac{\lambda(\epsilon + \tau M)}{\hbar^2\upsilon^2}\delta( z-z^\prime ),
&
\label{Eq1}\\
&
\left.\partial_z G_{ee|\tau k} =
q\, G_{ee|\tau k}\right|_{z=0},
\,
q=\frac{\lambda(\tau \varepsilon + M)}{\hbar \upsilon}  -\lambda k,
&
\label{BC1}
\end{eqnarray}
with $z=y/\lambda$, $z_k=-2\lambda\, k\, {\rm sgn} (eB)$, $\lambda=\sqrt{ c\hbar/2|eB| }$,  and
$a=\lambda^2(M^2 - \epsilon^2)/\hbar^2 \upsilon^2 - \,{\rm sgn} (eB)/2$.
The equations for $G_{hh|\tau k}$ are obtained from Eqs.~(\ref{Eq1}) and (\ref{BC1}) by replacement 
$\tau, k, B\to  -\tau, -k, -B$.
The detailed solution to Eqs.~(\ref{Eq1}) and (\ref{BC1}) has been given in Ref.~\cite{GT10b}. 
The outcome of these calculations is that the Green's function $G_{ee|\tau k}$ can be expressed in terms of the parabolic cylinder function $U( a , z)$ \cite{AS} as follows
\begin{eqnarray}
G_{ee|\tau k} = G^{\infty}_{ee|\tau k}(z,z^\prime) &-& C \frac{ \partial_{z_k}U( a , z_k )   + q U( a , z_k )  }
          { \partial_{z_k} U( a ,- z_k ) + q U( a , -z_k ) }
\nonumber\\
&\times & 
U( a , z - z_k )U( a , z^\prime - z_k ).
\label{G_ee}
\end{eqnarray}
The last term is the contribution of the edge, whereas
\begin{eqnarray}
G^\infty_{ee|\tau k} &=& C
\left[
\Theta(z - z^\prime)
U(a,z-z_k)U(a,-z^\prime+z_k)
\right.
\nonumber\\
&+&
\left.
\Theta(z^\prime -z)
U(a,z^\prime-z_k)U(a,-z+z_k)
\right],
\label{G_ee_bulk}\\
C &=& -\lambda( \epsilon + \tau M)\Gamma(a +1/2)/\sqrt{2\pi}\hbar^2\upsilon^2,
\label{C}
\end{eqnarray}
is the Green's function of the unbounded system (source term), where $\Gamma(a +1/2)$ is Euler's gamma function.
We then insert $G_{ee|\tau k}$ and $G_{hh|\tau k}$ into Eq. (\ref{G_diag}) and eliminate all the derivatives, using 
the recurrence relations~\cite{AS} for $U(a,z)$  and assuming, for concreteness, $eB>0$. 
As a result, the edge contribution takes the following form:
\begin{eqnarray}
\hat G_{\tau k}=
\frac{ \alpha(z,z^\prime) \left(
\begin{array}{cc}
1 & \beta(z^\prime)\\
\beta(z) & \beta(z)\beta(z^\prime)
\end{array}
\right) }{\epsilon - \tau M - \tau (\hbar v/\lambda) U(a, -z_k)/U(a+1,-z_k)  },
\label{G}
\end{eqnarray}
with functions $\alpha(z,z^\prime)$ and $\beta(z)$ given by
\begin{eqnarray}
&&
\alpha(z,z^\prime) = \frac{U(a, z-z_k) U(a, z^\prime-z_k) }{\lambda U(a, -z_k)U(a+1,-z_k)},
\label{alpha}\\
&&
\beta(z) = \frac{U(a, -z_k) U(a+1, z - z_k) }{U(a+1, -z_k) U(a, z - z_k)}.
\label{beta}
\end{eqnarray}
The new feature of solution (\ref{G}) is that it is valid for 
an arbitrary parameter $\hbar\upsilon/\lambda |M|$ 
which measures the magnetic field strength. 
Below we compare {\em weak}- and {\em strong}-field regimes defined by 
$\hbar\upsilon/\lambda |M|\leq 1$ and $\hbar\upsilon/\lambda |M|\gg 1$, respectively.

\subsection{Edge states: weak- and strong-field asymptotics}

\begin{figure}[t]
\begin{center}
\resizebox{0.75\columnwidth}{!}{%
  \includegraphics{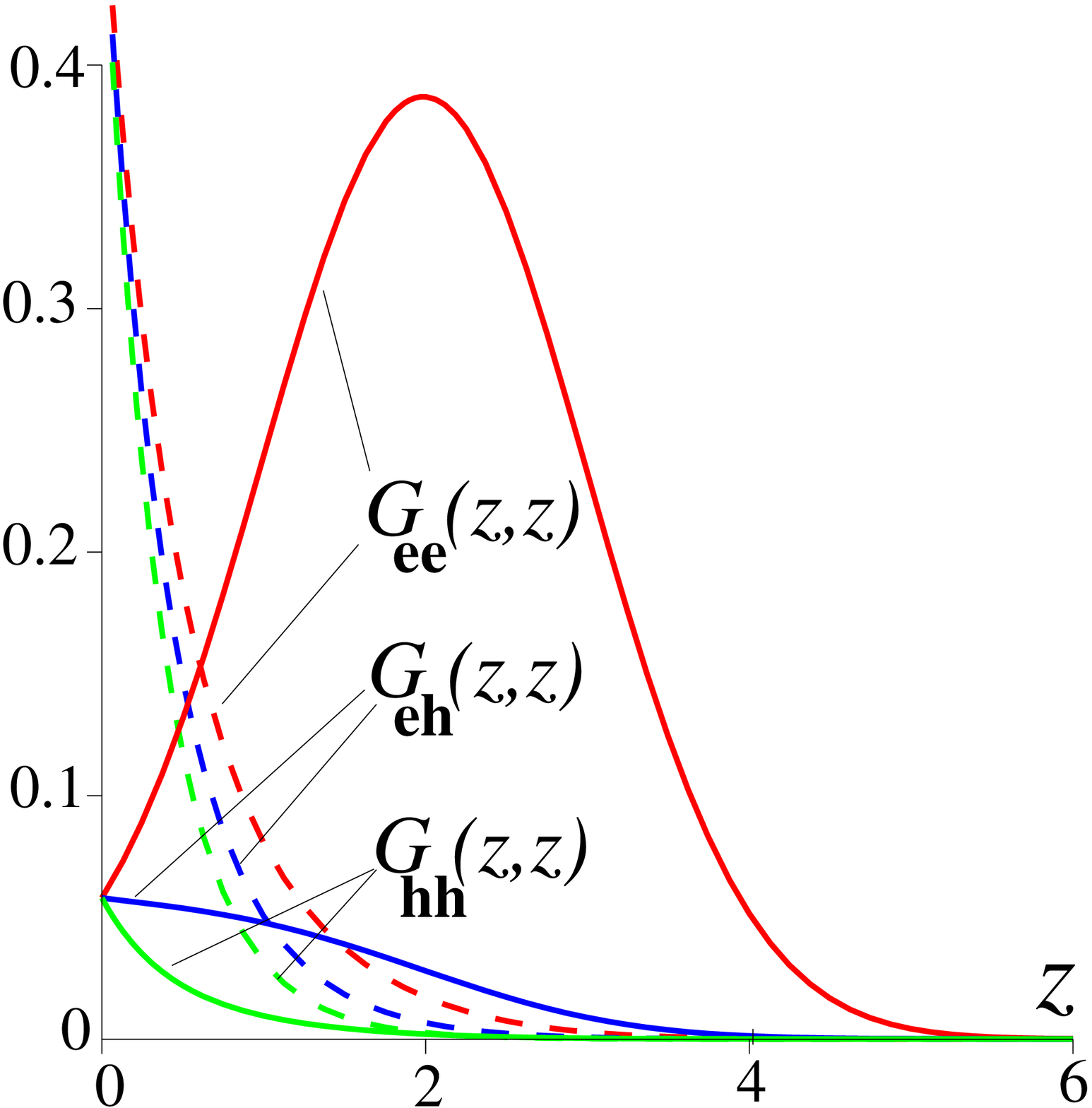}
}
\end{center}
\caption{
Spatial distribution of edge states in weak (dashed curves)
and strong (solid curves) magnetic fields at zero energy ($\epsilon=0$ and $M<0$) 
[see, Eqs.~(\ref{G_eh}), (\ref{G}) and text],
}
\label{G_z}
\end{figure}

The edge-state spectrum is given by the poles of Eq.~(\ref{G}), i.e. by the zeros of the equation
\begin{equation}
 \epsilon - \tau M - \tau (\hbar v/\lambda) U(a, -z_k)/U(a+1,-z_k)=0.
\label{Zeros}
\end{equation}
This equation describes the transition from the band insulator with $M>0$ to the QSH state with $M<0$ 
[cf. Figs.~\ref{Edge}(a) and (b)], which is observed at the critical QW thickness $\approx 6.3$ nm~\cite{Bernevig06,Koenig07}.
The QSH state has two gapless counter-propagating spin channels which are  
exponentially localized at the edge for weak magnetic fields, 
as seen from Eq.~(\ref{G}) and Fig.~\ref{G_z} where we use
the asymptotic formula $U(a,z)\approx \sqrt{\pi}/[ 2^{a/2 + 1/4} \Gamma(3/4 + a/2) ]{\rm e}^{-\sqrt{a} \, z}$
with $|a|\gg 1$~\cite{AS}, valid for low fields and energies $|\epsilon| < |M|$:
\begin{eqnarray}
{\hat G}_{\tau k}\approx\frac{(\sigma_0 + \sigma_x)\, \frac{ |M| }{\hbar\upsilon} 
{\rm e}^{ - |M|(y+y^\prime)/\hbar\upsilon }}{ \epsilon  -  \tau M\Theta(M) -  \tau \hbar \upsilon ( k  - k_B)  }, 
\, \frac{\hbar\upsilon}{\lambda  |M| }\ll 1.
\label{G_weak}
\end{eqnarray}
The subgap edge-state dispersion is linear: $\epsilon_{\tau k}= \tau \hbar \upsilon ( k  - k_B) $ for $M<0$. 
The magnetic field only shifts the zero-energy point $k_B= -eB \upsilon/(2c|M|)$ with no effect on transport.

As the magnetic field does not open a gap, the QSH state persists in strong fields $\hbar\upsilon/\lambda  |M| \gg 1$,
though the QSH channels are no longer localized at the edge [see, solid curves in Fig.~\ref{G_z}].
The electron function $G_{ee}(z,z)\propto \alpha(z,z)$ for $eB>0$ [or the hole one $G_{hh}(z,z)$ for $eB<0$]
behaves almost like the lowest-Landau-level bulk wave function
peaked at the center of oscillator $z_k$. The other functions 
are small at $z\sim z_k$. The strong-field asymptotic is obtained for $U(a,z)\approx U(-1/2,z)={\rm e}^{-z^2/4}$,
$U(a+1,z)\approx U(1/2,z)={\rm e}^{z^2/4}\sqrt{\pi/2}\,{\rm erfc}( z/\sqrt{2} )$, and $\beta\ll 1$
in Eqs.~(\ref{G}) and (\ref{alpha}):
\begin{eqnarray}
&&
{\hat G}_{\tau k}\approx  \frac{\sigma_0 + \sigma_z}{2}\,G_{\tau k},\qquad
G_{\tau k}(z,z^\prime)=\frac{\alpha(z,z^\prime)}{ \epsilon - \epsilon_{\tau k} },
\label{G_strong}\\
&&
\alpha(z,z^\prime)\approx \sqrt{ \frac{2}{\pi}  } \frac{ {\rm e}^{ -(z-z_k)^2/4 - (z^\prime-z_k)^2/4   } }
{\lambda\, {\rm erfc}( - z_k/\sqrt{2} )},
\label{alpha_strong}\\
&&
\epsilon_{\tau k}= \tau M + \tau \sqrt{ \frac{2}{\pi}  }\, \frac{\hbar\upsilon}{\lambda}\frac{ {\rm e}^{-z_k^2/2 } }
{ {\rm erfc}( - z_k/\sqrt{2} ) },\quad
\frac{\hbar\upsilon}{\lambda  |M| }\gg 1,\quad\quad
\label{E_strong}
\end{eqnarray}
where ${\rm erfc}( z )$ is the complementary error function. 
However, the most essential distinction of this regime is the {\em nonlinear} spectrum (\ref{E_strong}). 
Upon crossing the gap energy $\epsilon=|M|$ it changes from helical to chiral, 
as illustrated in Fig.~\ref{Edge}(c). Therefore, the QSH state transforms into  
a dissipationless $\nu =1$ QH state~\cite{Halperin82,MacDonald84}.  
The transition between the QSH and QH regimes can be accessed experimentally through the gate-voltage (i.e. energy) 
dependence of the longitudinal magnetoconductance. To analyze such a dependence (see next section)
we will need the group velocities, $\upsilon_\pm(\epsilon,B)$ and the center-of-oscillator coordinates, 
$z_\pm(\epsilon,B)$, which are obtained from Eq.~(\ref{E_strong}) linearized near given energy,
\begin{eqnarray}
\epsilon_{\tau k}\approx \epsilon - (\hbar\upsilon_\tau/2\lambda) (z_k - z_\tau).
\label{E_lin}
\end{eqnarray}
Here $z_{\tau}(\epsilon,B)$ is the solution of equation $\epsilon_{\tau k} =\epsilon$, 
which is related to the velocity by
\begin{eqnarray}
\upsilon_\tau(\epsilon,B)=2\lambda (\tau |M| +\epsilon)\, z_{\tau}(\epsilon,B)/\hbar.
\label{zv}
\end{eqnarray}
Consequently, in strong fields the edge state can be described by the one-dimensional Green's function,
\begin{eqnarray}
G_\tau (x,x^\prime)=\int \frac{dk}{2\pi}\,e^{i k (x-x^\prime)} \int_0^{\infty}dy\, G_{\tau k}(y,y),
\label{G_1D}
\end{eqnarray}
where $G_{\tau k}(y,y)$ is localized within $\lambda$ [see, Eq.~(\ref{alpha_strong})]. 
Using the linearized dispersion (\ref{E_lin}) we find
\begin{equation}
G_{\tau}(x,x^\prime)=\Theta([x-x^\prime] \tau)e^{ ik_\tau (\epsilon,B) (x-x^\prime) }/
i\hbar |\upsilon_\tau(\epsilon,B)|,
\label{G0}
\end{equation}
where the unit-step function $\Theta([x-x^\prime] \tau)$ accounts for the chirality and wave vector
$k_\tau=-z_\tau/2\lambda$ is related to the center-of-oscillator coordinate obtained numerically as discussed above.

\section{Backscattering of helical edge states and magnetoconductance} 

\begin{figure}[t]
\begin{center}
\resizebox{0.75\columnwidth}{!}{%
  \includegraphics{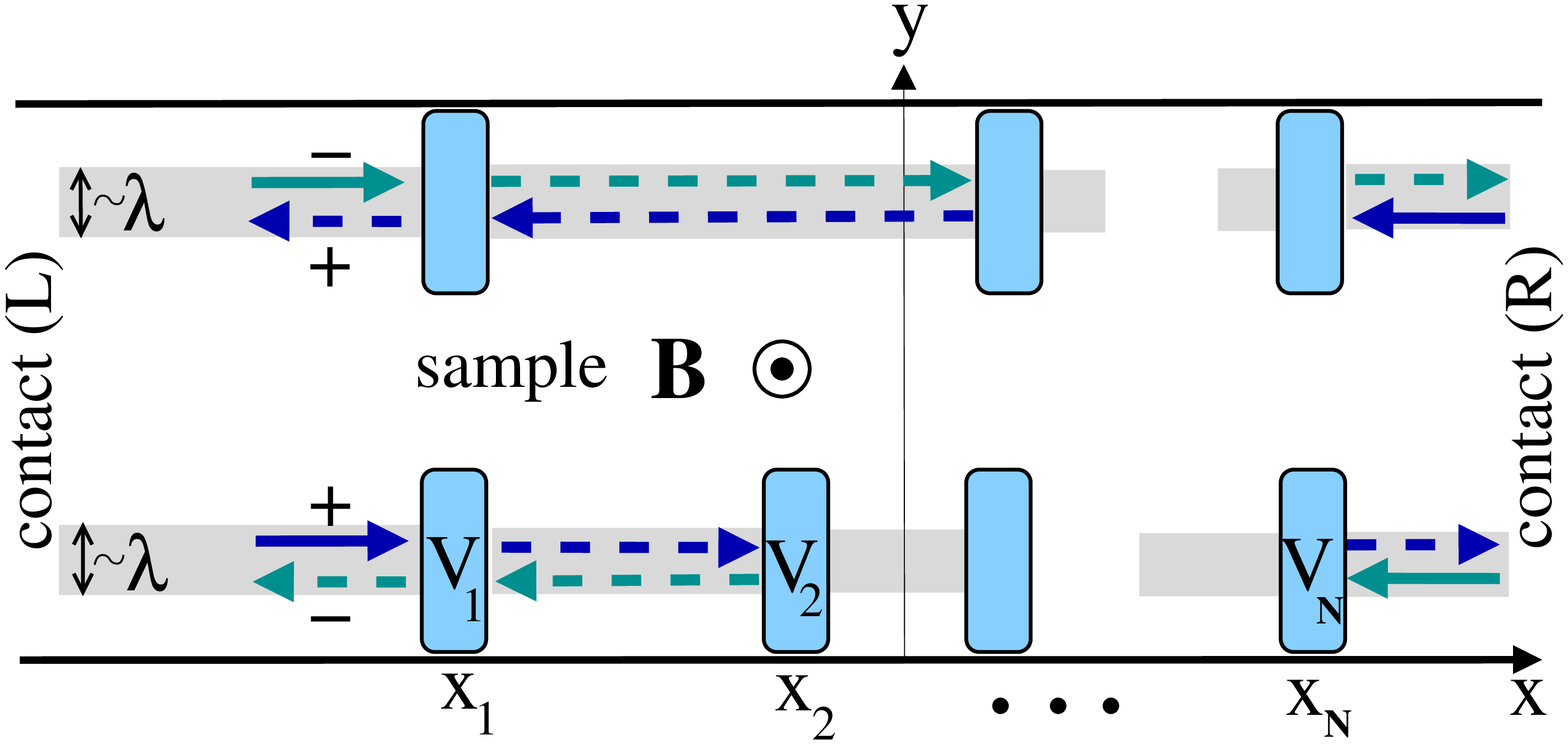}
}
\end{center}
\caption{
Two-terminal QSH system in a strong perpendicular magnetic field. 
Edge channels are localized within the magnetic length $\lambda$. 
Light blue regions schematically indicate backscattering centers (see also text). 
We assume that the current is carried by the right moving states (solid arrows $\pm$),
populated in contact L and equilibrating in contact R.}
\label{Geo}
\end{figure}

We now calculate the two-terminal magnetoconductance of a QSH system in  
the presence of backscattering centers, i.e. defects causing the scattering between the counter-propagating channels
[see, Fig.~\ref{Geo}]. Microscopically, such scattering  
can be mediated by interaction with electronic trap states which are likely to exist 
even in high quality samples~\cite{Roth09,EEI}.  
Since the edges are assumed decoupled, it is sufficient to do the calculation 
for one of them, e.g., for the lower edge in Fig.~\ref{Geo}. 
The scattering matrix for the edge can be decomposed in two parts:   
\begin{equation}
S= \frac{\sigma_0+\sigma_z}{2} \otimes {\check S}, \quad 
{\check S}=
\left(
  \begin{array}{cc}
   r^{-+}_{_{LL}}  & t^{--}_{_{ LR }} \\
   t^{++}_{_{ RL }}  &  r^{+-}_{_{ RR }} \\
  \end{array}
\right).
\label{S}
\end{equation}
The first factor $(\sigma_0+\sigma_z)/2$ is the projector on the electron QW subband
which has the non-vanishing wave function in the strong magnetic field [cf. Eq.~(\ref{G_strong})].
The second factor ${\check S}$ is the scattering matrix connecting right ("+")- and left ("-")- moving electron states with corresponding 
reflection $r$ and transmission $t$ amplitudes.
The conductance can be calculated as
\begin{equation}
 g=(e^2/h)\left| t^{++}_{_{RL}}\right|^2,
\label{g}
\end{equation}
using Fisher-Lee relation~\cite{FisherLee81}, 
\begin{equation}
t^{++}_{_{RL}}=i\hbar |\upsilon_+|{\cal G}_{++}(x\in R, x^\prime\in L),
\label{FisherLee}
\end{equation}
between $t^{++}_{_{RL}}$ and the diagonal element ${\cal G}_{++}(x,x^\prime)$ of the Green's function,  
\begin{eqnarray}
 {\check {\cal G} }(x,x^\prime) = \left(
  \begin{array}{cc}
   {\cal G}_{++}(x,x^\prime)  & {\cal G}_{+-}(x,x^\prime) \\
   {\cal G}_{-+}(x,x^\prime)  & {\cal G}_{--}(x,x^\prime) \\
  \end{array}
\right).
\label{G_matrix}
\end{eqnarray}
which is a matrix in space of the right- and left-movers and has {\em off-diagonal} elements due to backscattering. 
We model the backscatterers by the sum of $N$ potentials, localized at positions $x_n$ with  
non-zero matrix elements $V_n$ between the right- and left-moving states:   
\begin{equation}
{\check V}(x)=\sum\nolimits_{n=1..N} V_n \delta(x-x_n)\, \tau_x.
\label{V}
\end{equation}
Note that choosing the other off-diagonal matrix, $\tau_y$ does not change the final result. 

Potential ${\check V}(x)$ (\ref{V}) results in the Dyson equation 
\begin{eqnarray}
{\check {\cal G} }(x,x^\prime) = {\check G}(x,x^\prime) + 
\sum_{n=1..N}{\check G}(x,x_n)V_n\,\tau_x\,
{\check {\cal G} }(x_n,x^\prime),
\label{G_Dyson}
\end{eqnarray}
where 
\begin{eqnarray}
 {\check G}(x,x^\prime) = \left(
  \begin{array}{cc}
   G_{+}(x,x^\prime)  & 0 \\
   0  & G_{-}(x,x^\prime) \\
  \end{array}
\right)
\label{G0_matrix}
\end{eqnarray}
is the Green's function matrix in the absence of scattering [see, Eq. (\ref{G0})]. 
Using Eqs.~(\ref{V}), (\ref{G_Dyson}) and (\ref{G0_matrix}) we obtain the coupled equations for 
the diagonal ${\cal G}_{++}(x,x^\prime)$  and off-diagonal  ${\cal G}_{-+}(x,x^\prime)$ elements: 
\begin{eqnarray*}
&&
{\cal G}_{++}(x,x^\prime) = G_+(x,x^\prime) + 
\sum_{n} G_+(x,x_n)V_n {\cal G}_{-+}(x_n,x^\prime),
\label{G_Dyson_1}\\
&&
{\cal G}_{-+}(x,x^\prime) = \sum_{n} G_-(x,x_n)V_n {\cal G}_{++}(x_n,x^\prime).
\label{G_Dyson_2}
\end{eqnarray*}
Eliminating ${\cal G}_{-+}(x,x^\prime)$ yields a closed equation for ${\cal G}_{++}(x,x^\prime)$: 
\begin{eqnarray}
&&
{\cal G}_{++}(x,x^\prime) = G_+(x,x^\prime)+
\nonumber\\
&&
+
\sum_{n,m} G_+(x,x_n) V_n G_-(x_n,x_m) V_m {\cal G}_{++}(x_m,x^\prime).
\label{G_Dyson_++}
\end{eqnarray}
With known unperturbed function $G_\tau$ and for not large $N$, 
we solve this equation and calculate $g$.  
Let us look first at the particular cases $N=1,2$ and $3$:
\begin{eqnarray}
g=\frac{e^2}{h}
\left( 1+\frac{V_1^2}{\hbar^2|\upsilon_+\upsilon_-| } \right)^{-2},
\label{g1}
\end{eqnarray}
\begin{eqnarray}
g=\frac{e^2}{h}
\left| 1 + \frac{V^2_1+V^2_2+V_1V_2 e^{ iQd_{12} } }{\hbar^2|\upsilon_+\upsilon_-| }
         + \frac{ V^2_1V^2_2 }{\hbar^4\upsilon^2_+\upsilon^2_- }
\right|^{-2},
\label{g2}
\end{eqnarray}
\begin{eqnarray}
&&
g=\frac{e^2}{h}
\left|1 + \frac{ V_1^2V_2^2V_3^2 }{\hbar^6|\upsilon_+\upsilon_-|^3} \right.
\label{g3}\\
&&
+
\frac{
\sum_{n=1}^3V^2_n+V_1V_2 e^{iQd_{12}}  + V_1V_3 e^{iQd_{13}} +V_2V_3 e^{iQd_{23}} 
}
{\hbar^2|\upsilon_+\upsilon_-| }
\nonumber\\         
&&
\left.
+ \frac{ V_1^2V_2^2 + V_1^2V_3^2 + V_2^2V_3^2 + V_1^2V_2V_3 e^{ iQd_{23} } }{\hbar^4\upsilon^2_+\upsilon^2_-} 
\right|^{-2},
\nonumber
\end{eqnarray}
where $Q=k_+ - k_-$ and $d_{nm}=x_m-x_n$.

Clearly, for arbitrary $N$ the conductance contains 
the cross product $V_1^2\cdots V_N^2/|\upsilon_+\upsilon_-|^N$ arising from the simultaneous scattering 
from $N$ potentials. This is the most divergent term when one of the velocities $\upsilon_\pm$ 
vanishes near the band gap, $|\epsilon|\to E_g=|M|$  (e.g. $\upsilon_-\to 0$ in Fig.~\ref{Edge}(c)). 
Such strong enhancement of the backscattering leads to the suppressed conductance, 
\begin{eqnarray}
g\approx (e^2/h)\times\hbar^{4N}|\upsilon_+\upsilon_-|^{2N}/(V_1\cdots V_N)^4\ll e^2/h. 
\label{gN}
\end{eqnarray}
Using Eq.~(\ref{zv}) we obtain the qualitative energy and field dependence of the conductance discussed earlier 
in the introduction [see, Eq.~(\ref{g_as})].

\begin{figure}[t]
\begin{center}
\resizebox{0.8\columnwidth}{!}{%
  \includegraphics{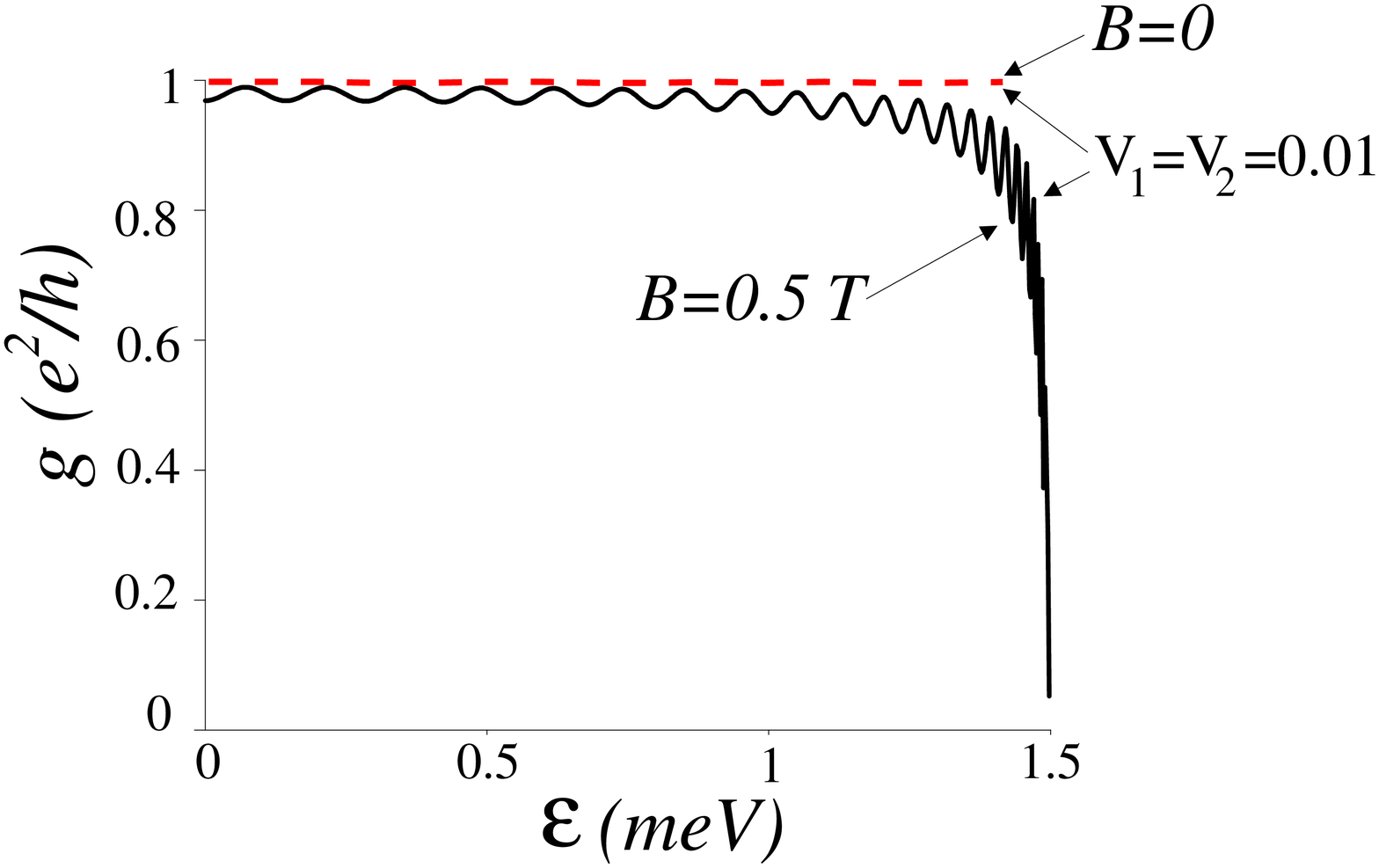}
}
\end{center}
\caption{ Conductance [see, Eq.~(\ref{g2})] versus energy below band gap $|M|=1.5$ meV. 
Scattering potential strengths $V_{1,2}$ are in units of meV$\cdot\mu$m; $d_{12}=3 \mu$m.}
\label{g_E}
\end{figure}

\begin{figure}[t]
\begin{center}
\resizebox{1\columnwidth}{!}{%
  \includegraphics{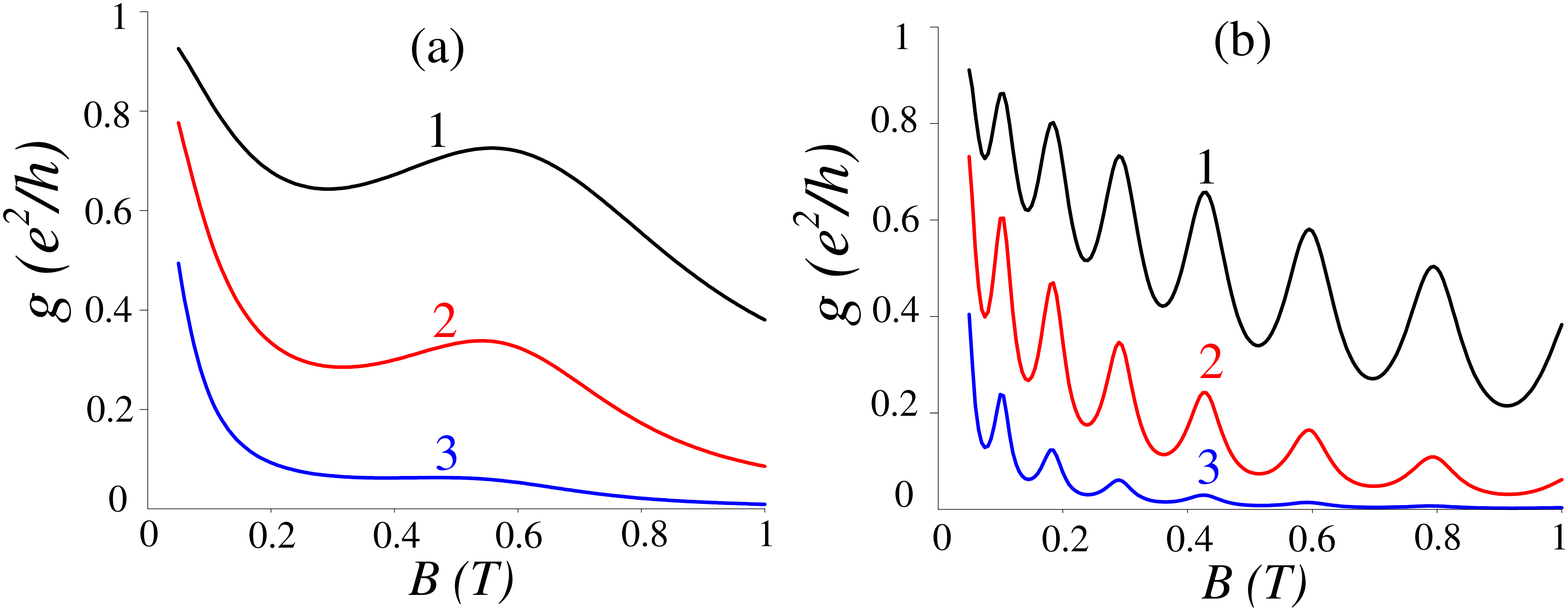}
}
\end{center}
\caption{ Conductance [see, Eq.~(\ref{g2})] vs. magnetic field for different energies within the band gap: 
(a) $\epsilon=0.2$ meV and (b) $\epsilon=1$ meV. Labels 1, 2 and 3 refer to   
scattering potential strengths $V_1=V_2=0.06$, $V_1=V_2=1$ and $V_1=V_2=0.15$ meV$\cdot\mu$m, respectively; $d_{12}=3 \mu$m.}
\label{g_B}
\end{figure}

In a wider range of energies and fields the typical behavior of the conductance can be understood 
from Eq.~(\ref{g2}) assuming two backscattering centers on the edge.   
First of all, it is easy to verify that Eq.~(\ref{g2}) is valid not only for strong fields,
but also in the weak-field case where the unperturbed Green's function is given by Eq.~(\ref{G_weak}).
Since the weak-field spectrum is linear $\epsilon_{\tau k}= \tau \hbar \upsilon ( k  + k_B) $, 
we have $\upsilon_+=-\upsilon_-=\upsilon$, $k_{\pm }= - k_B \pm \epsilon/\hbar\upsilon $ and $Q=k_+-k_-=2\epsilon/\hbar\upsilon$. 
Therefore, $g$ is independent of the magnetic field 
and for $V_{1,2}\ll \hbar\upsilon$ is almost independent of energy (see dashed curve in Fig.~\ref{g_E}).
Thus weak channel mixing is hardly detectable for small $B$.
Also, since we focus on the quasi-ballistic systems with just a few backscatteres, there is no gap opening in this case 
[this conclusion may no longer be true for systems with a sizable macroscopic number of such defects].
In contrast, in strong magnetic fields, scattering of the same strength is sufficient to 
suppress the conductance, $g$ near the band gap where the QSH-QH transition occurs  
[cf. dashed and solid curves for $V_{1,2}=0.01$ meV$\cdot\mu$m near $|M|=1.5$ meV in Fig.~\ref{g_E}].
Figs~\ref{g_E} and ~\ref{g_B} also show that the conductance suppression 
is accompanied by Fabry-Perot-type oscillations due to interference of the counter-propagating channels 
which acquire the energy- and field-dependent 
phase difference $(k_+ - k_-)d_{12}$, in scattering between the defects.

In our model the upper magnetic field limit lies in the range of a few Tesla. 
This estimate is based on Refs.~\cite{Koenig08,Schmidt09} predicting another ($B$-field induced) 
QSH-QH transition due to a quadratic correction ${\cal B}{\bf k}^2$ to the mass term in Eq.~(\ref{Heff}). 
The smallness of the parameter ${\cal B}|M|/2\hbar^2\upsilon^2\ll 1$~\cite{Koenig08,Schmidt09} 
allows us to neglect such ${\bf k}^2$ term and to meet, at the same time, 
the strong field condition $\hbar\upsilon/\lambda |M|> 1$. 

\section{Conclusions} 

We have studied helical edge channels and their longitudinal conductance in a two-dimensional topological insulator subject to strong quantizing magnetic fields. The helical edge channels consist of a pair of counter-propagating states that exist within a bulk band gap of the topological insulator. We have shown that, albeit persistent in a strong magnetic field, the counter-propagating states acquire different group velocities. In particular, at the band gap the helical spectrum turns chiral: one of the edge states merges with the dispersionless bulk Landau level, 
whereas the other remains propagating, which corresponds to the onset of the $\nu=1$ quantum Hall state. 
Due to the drastically different group velocities of the helical modes,  
the longitudinal conductance is very sensitive to backscattering that couples the counter-propagating channels. 
We have found that in the presence of backscattering the longitudinal conductance rapidly decreases as a function of both Fermi energy and magnetic field. 
It shows a power-law magnetic field dependence $B^{-2N}$, determined by the number $N$ of backscattering centers on the edge. 
This suggests a simple way to detect such defects in ballistic QSH devices 
using standard magnetoresistance measurements. 

Our findings may have implications for the studies of other related phenomena in HgTe quantum structures (see e.g. Refs.~\cite{Yang08,Chang11}) 
including hybrid structures such as topological insulator/superconductor junctions.  
In hybrid structures consisting of conventional two-dimensional semiconductors and supersonductors the magnetotransport 
is strongly influenced by Andreev reflection~\cite{Hoppe00,GT04,GT05a,GT05b,Eroms05,Batov07,Rohlfing09}, 
whereby Cooper pairs are transferred between superconducting and normal regions. 
Since topological insulators differ markedly from conventional semiconductors, 
our analysis may help to understand Anreev magnetotransport through helical edge channels.

We thank S. C. Zhang, Q.L. Qi, J. Maciejko, A. Novik, H. Buhmann, L.W. Molenkamp, B. Trauzettel and A. H. MacDonald 
for helpful discussions. This work was funded through DFG  Grant HA5893/1-1.






\begin{thebibliography}{99}


\bibitem{Kane05}
C. L. Kane and E. J. Mele, Phys. Rev. Lett. {\bf 95}, 226801 (2005).

\bibitem{Bernevig06}
B. A. Bernevig and T. L. Hughes and S. C. Zhang,
Science {\bf 314}, 1757 (2006).

\bibitem{Koenig07}
M. K{\"o}nig, S. Wiedmann, C. Br{\"u}ne, A. Roth, H. Buhmann, L. W. Molenkamp, X.-L. Qi and S.-C. Zhang,
Science {\bf 318}, 766 (2007).


\bibitem{Fu07}
L. Fu and C. L. Kane, Phys. Rev. B {\bf 76}, 045302 (2007).  

\bibitem{Moore07}
J. E. Moore and L. Balents, Phys. Rev. B {\bf 75}, 121306(R) (2007).

\bibitem{Hsieh08}
D. Hsieh, D. Qian, L. Wray, Y. Xia, Y. S. Hor, R. J. Cava, and  M. Z. Hasan,
Nature {\bf 452}, 970 (2008).

\bibitem{Xia09}
Y. Xia, D. Qian, D. Hsieh, L. Wray, A. Pal, H. Lin, A. Bansil, D. Grauer, Y. S. Hor, R. J. Cava, and  M. Z. Hasan,
Nature Phys. {\bf 5}, 398 (2009).

\bibitem{Chen09}
Y. L. Chen, J. G. Analytis, J.-H. Chu, Z. K. Liu, S.-K. Mo, X. L. Qi, H. J. Zhang, D. H. Lu, X. Dai, Z. Fang, S. C. Zhang, I. R. Fisher, Z. Hussain, and  Z.-X. Shen,
Science {\bf 325}, 178 (2009).

\bibitem{Koenig08}
M. K\"onig, H. Buhmann, L. W. Molenkamp, T. Hughes, C.-X. Liu, X.-L. Qi, and S.-C. Zhang, 
J. Phys. Soc. Jpn. {\bf 77}, 031007 (2008) and references therein. 

\bibitem{Hasan10}
M. Z. Hasan and C. L. Kane, Rev. Mod. Phys. {\bf 82}, 3045 (2010) and references therein. 

\bibitem{Qi10} 
X.-L. Qi and S.-C. Zhang, Rev. Mod. Phys. {\bf 83}, 1057 (2011) and references therein.


\bibitem{Qi08}
X.-L. Qi, T. L. Hughes, and S.-C. Zhang, Phys. Rev. B {\bf 78}, 195424 (2008).

\bibitem{Essin09}
A. M. Essin, J. E. Moore, and D. Vanderbilt, Phys. Rev. Lett. {\bf 102}, 146805 (2009).

\bibitem{Tse10a}
W.-K. Tse and A. H. MacDonald, Phys. Rev. Lett. {\bf 105}, 057401 (2010).

\bibitem{Tse10b}
W.-K. Tse and A. H. MacDonald, Phys. Rev. B {\bf 82}, 161104(R) (2010).

\bibitem{Bruene11}
C. Br\"une, C. X. Liu, E. G. Novik, E. M. Hankiewicz, H. Buhmann, Y. L. Chen. X. L. Qi, Z. X. Shen, S. C. Zhang, and L. W. Molenkamp, 
Phys. Rev. Lett. {\bf 106}, 126803 (2011).

\bibitem{Maciejko10a}
J. Maciejko, X.-L. Qi, H. D. Drew, and S.-C. Zhang, Phys. Rev. Lett. {\bf 105}, 166803 (2010).

\bibitem{Garate10}
I. Garate and M. Franz, Phys. Rev. Lett. {\bf 104}, 146802 (2010).

\bibitem{GT10c}
G. Tkachov and E. M. Hankiewicz, Phys. Rev. B {\bf 84}, 035405 (2011). 

\bibitem{Wilczek87}
F. Wilczek, Phys. Rev. Lett. {\bf 58}, 1799 (1987).


\bibitem{Roth09}
A. Roth, C. Br{\"u}ne, H. Buhmann, L. W. Molenkamp, J. Maciejko, X.-L. Qi, and S.-C. Zhang,
Science {\bf 325}, 294 (2009).


\bibitem{Buettner11}
B. B\"uttner, C. X. Liu, G. Tkachov, E. G. Novik, C. Br\"une, H. Buhmann, E. M. Hankiewicz, P. Recher, B. Trauzettel, S. C. Zhang and L. W. Molenkamp, 
Nature Phys. {\bf 7}, 418 (2011).

\bibitem{GT11a}
G. Tkachov, C. Thienel, V. Pinneker, B. B\"uttner, C. Br\"une, H. Buhmann, L. W. Molenkamp, and E. M. Hankiewicz, 
Phys. Rev. Lett. {\bf 106}, 076802 (2011).  

\bibitem{GT11b}
G. Tkachov and E. M. Hankiewicz, Phys. Rev. B {\bf 84}, 035444 (2011). 

\bibitem{Halperin82}
B.~I. Halperin, Phys. Rev. B {\bf 25}, 2185 (1982).

\bibitem{MacDonald84}
A.~H.  MacDonald and P. Streda, Phys. Rev. B {\bf 29}, 1616 (1984).


\bibitem{Maciejko10b}
J. Maciejko, X.-L. Qi, and S.-C. Zhang, Phys. Rev. B {\bf 82}, 155310 (2010). 

\bibitem{GT10a}
G. Tkachov and E. M. Hankiewicz, Phys. Rev. Lett. {\bf 104}, 166803 (2010).

\bibitem{Rothe10}
D. G. Rothe, R. W. Reinthaler, C.-X. Liu, L. W. Molenkamp, S.-C. Zhang, and E. M. Hankiewicz,
New J. Phys. {\bf 12}, 065012  (2010).

\bibitem{GT07}
G. Tkachov, Phys. Rev. B {\bf 76}, 235409 (2007).

\bibitem{GT09a}
G. Tkachov, Phys. Rev. B {\bf 79}, 045429 (2009).

\bibitem{GT09b} 
G. Tkachov and M. Hentschel, Phys. Rev. B {\bf 79}, 195422 (2009).

\bibitem{GT09c}
G. Tkachov and M. Hentschel, Eur. Phys. J. B {\bf 69}, 499 (2009).  

\bibitem{TB_GF}
for tight-binding Green's function methods, see,
P. Burset, A. Levy Yeyati, and A. Martin-Rodero, Phys. Rev. B {\bf 77}, 205425 (2008);
P. Burset, W. Herrera, and A. Levy Yeyati, Phys. Rev. B {\bf 80}, 041402 (2009); 
W. Herrera, P. Burset,  and A. Levy Yeyati, J. Phys.: Condens. Matter {\bf 22}, 275304 (2010).  


\bibitem{Berry87}
M. V. Berry and R. J. Mondragon, Proc. R. Soc. Lond. A {\bf 412}, 53 (1987).

\bibitem{Volkov85}
The 3D analogue of this problem was discussed by 
B. A. Volkov and O. A. Pankratov, Pis'ma Zh. Eksp. Teor. Fiz. {\bf 42}, 145 (1985) [JETP Lett. {\bf 42}, 178 (1985)].

\bibitem{GT10b}
G. Tkachov and E. M. Hankiewicz, Phys. Rev. B {\bf 83}, 155412 (2011).

\bibitem{AS}
M. Abramowitz and I. Stegun,  
{\em Handbook of Mathematical Functions with Formulas, Graphs, and Mathematical Tables} 
(National Bureau of Standards, Washington, D.C., 1964). 

\bibitem{EEI}
The electron-electron interaction could provide another mechanism of the backscattering of the helical edge states, 
which however requires separate discussion.

\bibitem{FisherLee81} 
D.~S. Fisher and P.~A. Lee, Phys. Rev. B {\bf 23}, 6851 (1981).

\bibitem{Schmidt09} 
M. J. Schmidt, E. G. Novik, M. Kindermann, and B. Trauzettel,
Phys. Rev. B {\bf 79}, 241306(R) (2009).

\bibitem{Yang08}
W. Yang, K. Chang and  S.-C. Zhang, Phys. Rev. Lett. {\bf 100}, 056602 (2008). 

\bibitem{Chang11}
K. Chang and W.-K. Lou, Phys. Rev. Lett. {\bf 106}, 206802 (2011).


\bibitem{Hoppe00}
H. Hoppe, U. Z\"ulicke, and G. Sch\"on, Phys. Rev. Lett. {\bf 84}, 1804 (2000).

\bibitem{GT04}
G.~Tkachov and V.~I.~Fal'ko, Phys.~Rev.~B {\bf 69}, 092503 (2004).

\bibitem{GT05a}
G.~Tkachov and K.~Richter, Phys.~Rev.~B {\bf 71}, 094517 (2005). 

\bibitem{GT05b}
G. Tkachov, Physica C {\bf 417}, 127 (2005).

\bibitem{Eroms05}
J. Eroms,  D. Weiss, J. De Boeck, G. Borghs, and U. Z\"ulicke, Phys.~Rev.~Lett.~{\bf 95}, 107001 (2005).

\bibitem{Batov07}
I. E. Batov, Th. Sch\"apers, N. M. Chtchelkatchev, H. Hardtdegen, and A. V. Ustinov, Physical Review B {\bf 76}, 115313 (2007).

\bibitem{Rohlfing09}
F. Rohlfing, G. Tkachov, F. Otto, K. Richter, D. Weiss, G. Borghs, and C. Strunk,
Phys. Rev. B {\bf 80}, 220507 (R) (2009).

\end{thebibliography}







\end{document}